\documentclass[aps,prl,twocolumn,groupedaddress]{revtex4}
\usepackage{graphicx,latexsym}

\begin{document}

\maketitle

{\Large \bf Experimental phase diagram of moving vortices}
\\ \\
{\bf M. Hilke, S. Reid, R. Gagnon, and Z. Altounian}
\\ \\
{\em Dpt. of Physics, McGill University, Montr\'eal, Canada H3A
2T8}

{\bf In the mixed state of type II superconductors, vortices
penetrate the sample and form a correlated system due to the
screening of supercurrents around them. Interestingly, we can
study this correlated system as a function of density and driving
force. The density, for instance, is controlled by the magnetic
field, $B$, whereas a current density $j$ acts as a driving force
$F=j\times B$ on all vortices. The free motion of vortices is
inhibited by the presence of an underlying potential, which tends
to pin the vortices. Hence, to minimize the pinning strength we
studied a superconducting glass in which the depinning current is
10 to 1000 times smaller than in previous studies, which enables
us to map out the complete phase diagram in this new regime. The
diagram is obtained as a function of $B$, {\em driving current}
and {\em temperature} and led a remarkable set of new results,
which includes a huge peak effect, an additional reentrant
depinning phase and a driving force induced pinning phase.}

The Peak effect (PE), which is one of the most intriguing
consequences of the motion of vortices in superconductors, is a
peak in the critical current as a function of $B$ and typically
occurs below, but near, the critical field ($B_{c2}$) in some
strong type II conventional superconductors
\cite{belincourt,rosenblum,pippard,kes,higginsexp}. In high
temperature superconductors the $PE$ is also observed but usually
appears well below $B_{c2}$ \cite{hitcexp} as inferred from
magneto-transport measurements. The $PE$ can lead to a reentrant
superconducting phase as a function of $B$ or $T$ when a current
slightly below the maximum critical current is applied. Above the
critical current, vortices experience a force strong enough to be
depinned and they start moving, which leads to a non-zero
electrical field $(E=\bar{v}\times B)$ and hence to a non-zero
resistance, where $\bar{v}$ is the average vortex velocity. At the
$PE$ the vortices are pinned again, which decreases the resistance
and enhances the critical current. While the exact mechanism of
the $PE$ is not fully understood, recent experiments on $Nb$ have
correlated the $PE$ with the disappearance of neutron scattering
Bragg peaks from the vortex structure. This was attributed to the
transition from a long-range ordered phase to a short-ranged
disordered phase, which in turn is pinned more efficiently
\cite{proorderdisorder,troyanovski}. This transition can be
induced either by increasing disorder or equivalently by
increasing $B$, since the disorder potential couples to all
vortices, hence effectively increasing with $B$. However, recent
experiments on $NbSe_2$ have questioned the amorphous nature of
the phase in the $PE$ regime \cite{conorderdisorder}.

In this work we investigate the vortex dynamics in a system with
the weakest possible pinning potential. It is well known that
amorphous superconductors have a lower critical current than
crystals because of the absence of long-range order, which implies
weaker collective pinning. The remaining pinning is then mainly
governed by impurities. We therefore used high purity $Fe-Ni-Zr$
based superconducting glasses, which have a similar critical
temperature ($T_c$) and $B_{c2}$ as the widely studied crystalline
$2H-NbSe_2$ system \cite{higginsexp} and others
\cite{rosenblum,pippard,hitcexp,troyanovski,proorderdisorder,
conorderdisorder}, but our amorphous samples exhibit a critical
current density, ($J_c\leq 0.4 A/cm^2$), just below the $PE$t,
which is typically 100 to 1000 times smaller. In comparison to
earlier studies on the $PE$ in amorphous films, our samples still
have a $J_c$ 10 times smaller
\cite{belincourt,kes,kes2,wordenweber}, which reflects the high
level of purity of our samples.

We obtained these high purity superconducting glasses by melt
spinning \cite{altounian} $Fe_xNi_{1-x}Zr_2$ with different values
of x. Although we only present results for x=0.3, qualitatively
very similar features were obtained for x=0, 0.1 and 0.2, which
suggests that these effects are inherent to these types of
materials and do not depend critically on composition. The
transition temperature ($T_c$) for x=0.3 is $2.30\pm 0.02$K as
extracted from three different samples. The sharpness of the
transition region is quite remarkable, with a transition
temperature width (10-90\% value) ranging between 5 and 20mK for
all three samples, indicative of very homogenous samples. The
absence of crystallinity is confirmed by the absence of Bragg
peaks in X-ray diffraction. The samples have the following typical
sizes ({\em thickness}=$21\mu$m, {\em width}=1.15mm, and {\em
length} between indium contacts=8mm). Using standard expressions
for superconductors in the dirty limit \cite{kes}, we can estimate
the different length scales in our system. The zero temperature
penetration depth is $\lambda=1.05\times
10^{-3}(\rho_N/T_C)^{1/2}\simeq 0.9\mu m$, the BCS coherence
length is $\xi\simeq 7.3nm$ (or 8.3nm using GL), the
Ginzburg-Landau (GL) parameter $\kappa\simeq 76$ and $B_{c1}\simeq
29$mT. Relevant experimental quantities are $T_c=(2.30\pm 0.02)$K,
$B_{c2}(T=0)=(4.8\pm 0.1)$T, $\rho_N=(1.7\pm0.2)\mu\Omega$m
(normal resistivity) and
$\left.\frac{dB_{c2}}{dT}\right|_{T_c}=(-2.7\pm0.2)$T/K. These
numbers are typical for a strong type II low-$T_c$ material.

Since moving vortices give rise to a non-zero resistance in a
superconductor, we measured the $B$-dependence of the
magneto-resistance for different currents to probe the dynamics of
the vortices. The results are presented in figure 1. The most
striking feature is the $B$-induced reentrant superconducting
phase for currents above 0.07mA. Such a strong reentrant
superconductor was never observed before in any amorphous
superconductor. However, because we observed a similarly strong
reentrant behavior for different concentrations of $Fe$ ($x$=0,
0.1, and 0.2), we believe that this effect is very general for
high purity and weakly pinned amorphous superconductors. Indeed,
when comparing our results to previous studies on the $PE$ in
amorphous superconductors \cite{belincourt,kes,kes2,wordenweber},
the most notable difference is the critical current, which is a
measure of the pinning strength and is at least 10 times smaller
in our samples, indicating that weak pinning enhances the
reentrant behavior. The reentrant behavior is the most dramatic
signature of the $PE$, which disappears below 0.07mA and implies a
peak in the critical current in the region of the reentrant
superconductor.

\begin{figure}[h]
\begin{center}
\vspace*{0cm}
\includegraphics[scale=0.40]{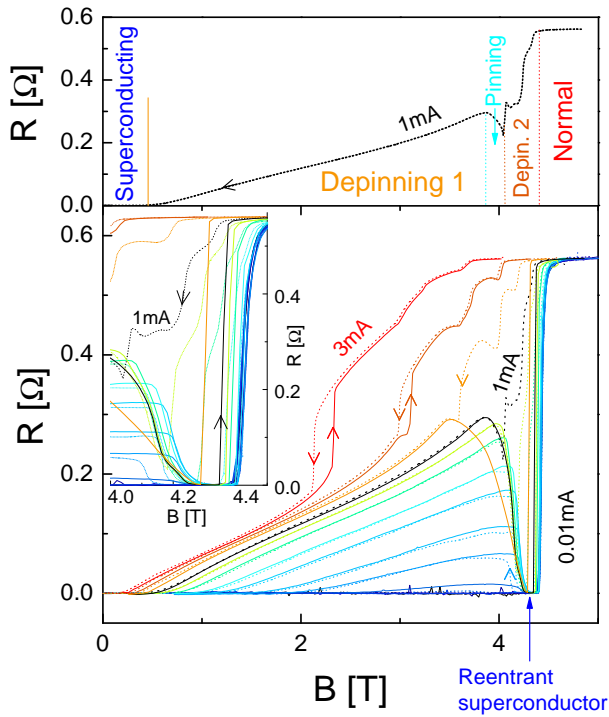}
\vspace*{-1cm} \caption{Lower curves: Resistances as function of
$B$ for up and down $B$-sweeps. The left curves expand the high
field region. The different curves are for different currents,
ranging from 0.01mA to 3mA (0.01, 0.02, 0.04, 0.07, 0.1, 0.15,
0.2, 0.3, 0.4, 0.6 0.8 1. 1.5, 2.1, 3)mA. The upper curve, which
is the resistance for the down sweep with 1mA, illustrates how we
determine the different transition points. The resistances were
measured using an AC resistance bridge at 17Hz and $T=450$mK}
\end{center}
\end{figure}

We map-out the phase diagram of the moving vortices using the
dissipative transport in figure 1 and illustrated by the top
figure with the help of the 1mA down-sweep curve. The results of
the diagram are presented in figure 2. At low fields we have the
first depinning transition defined when the resistance exceeds
0.5m$\Omega$, which is our experimental resolution. The choice of
this cut-off is not critical since close to the depinning
transition the dependence of the resistance on $B$ is stronger
than exponential. When increasing $B$ further, the system becomes
more and more resistive for currents above 0.07 mA. In this
region, which we denote {\em depinning 1}, the vortices start
moving. At even higher $B$, the moving vortices are pinned again
in the {\em pinning} region, which leads to the reentrant
superconducting phase inside this {\em pinning} region. The
pinning transition is defined when $dR/dB=0$. Inside the {\em
pinning} region, the resistance eventually vanishes within our
experimental resolution (0.5m$\Omega$). Finally, at high enough
$B$ we cross $B_{c2}$ and the system becomes {\em normal}.
$B_{c2}$ is defined as the point of strongest negative curvature
just before reaching the normal state. At higher driving currents
we observe another region, labelled {\em depinning 2}, which
corresponds to a sudden increase of resistance after the {\em
pinning} or {\em depinning 1} regions, indicative of another
depinning transition, before reaching $B_{c2}$, which is defined
as the point of highest positive curvature.

\begin{figure}[h]
\begin{center}
\vspace*{0cm}
\includegraphics[scale=0.5]{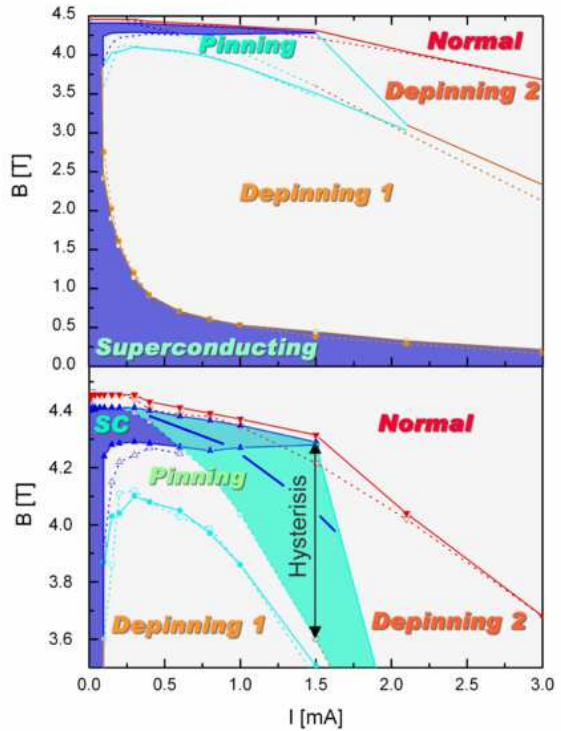}
\vspace*{-0cm} \caption{Phase diagram of vortex dynamics as
extracted from fig. 1. The solid lines are obtained when sweeping
$B$ up and the dotted lines when $B$ is swept down. The lower half
is an enlargement of the high $B$ region. Overall, the diagram is
consistent with three distinct phases ({\em depinning 1, pinning}
and {\em depinning 2}) connected by a strongly hysteretic triple
point at the end of the thick dashed line.}
\end{center}
\end{figure}

A striking feature of the phase diagram is the huge $PE$ at 4.3T.
Indeed, the critical current, which delimits the superconductor
(shaded in dark blue in figure 2) from the non-zero resistance
depinned phases, jumps from 0.07mA to 1.5mA close to 4.3T. This
$PE$ is associated with the {\em depinning 1} - to - {\em pinning}
transition of the vortices.

Quite generally, the {\em depinning 1} region is characterized by
weakly pinned vortices, which move beyond a $B$-dependent
activated threshold. Giamarchi and co-workers \cite{giamarchiBG}
argued that this phase is a long-range ordered moving Bragg glass
(MBG), consistent with a recent experiment showing algebraic
neutron scattering Bragg peaks in this region with measurements
performed on $(K,Ba)BiO_3$ \cite{giamarchiBGexp}. Our results
support this picture since the dependence of the voltage (V) on
current (I) below the $PE$ for $B$-fields between 0.25 and 3.75T,
(as illustrated in the inset of figure 3), is well fitted by the
activated creep expression $V\sim e^{-U/T\sqrt{I}}$. $U$ reflects
the pinning strength and the expression was derived directly from
the equation of motion \cite{creepIV} and assumes a long-range
ordered phase, such as a Bragg glass. The long-range nature of
this phase can be inferred from the geometry dependent transition
point. Indeed, when tilting the field so that $B$ is aligned along
the wide axis of the sample (instead of perpendicular to it), the
transition is shifted upwards between 1.5T and 2.5T depending on
the current. Since the sample is amorphous and 1mm wide but only
20$\mu$m thick this implies long range order above 20$\mu$m. This
transition is the only one, which is significantly affected by the
tilted field below 1.5mA, suggesting that the remaining phases
have short-ranged order.

\begin{figure}[h]
\begin{center}
\vspace*{1.5cm}
\includegraphics[scale=0.35]{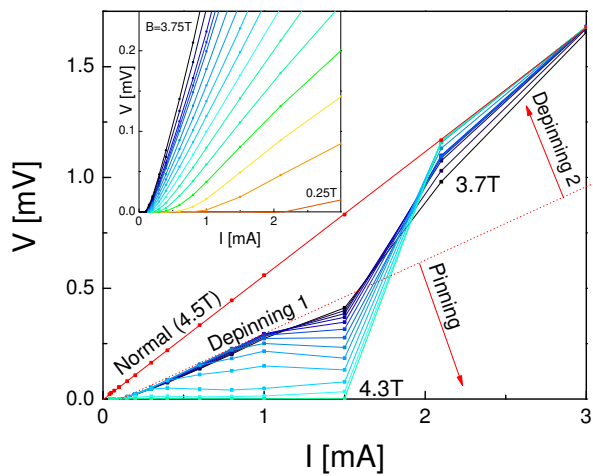}
\vspace*{-3cm} \caption{Dependence of the resistance as a function
of excitation voltage for increasing field. Inset: dependence of
the voltage on driving current for decreasing field.}
\end{center}
\end{figure}

At higher $B$ the pinning increases in the $PE$ region. In this
region the inter-vortex distance is close to $\xi$, which is the
size of the vortex core and implies strong inter-vortex
correlations. The exact nature of this pinning transition is still
under debate, but is commonly associated with an order-disorder
transition \cite{giamarchirev}, the melting of the vortex
structure \cite{larkin} or the anomalous friction close to
$B_{c2}$ \cite{nissila}. Experimentally, the reentrant
superconductor is nonetheless the most spectacular demonstration
for the existence of a {\em depinning} - to - {\em pinning}
transition. In figure 2, the reentrant superconductor is shaded in
dark blue, which is included in the {\em pinning} region delimited
by the light blue line. We mapped out the {\em pinning} regions
for both $B$-sweep directions up (solid lines) and down (dotted
lines) and observed an increasing hysteresis as a function of
driving current, which is the region shaded in light blue. The
region of hysteresis does {\em not} depend on the sweeping rate
and is stable once reached. The left inset of figure 1 shows
details of the region of hysteresis and the $B$-sweep direction is
labelled with corresponding arrows. The hysteresis is particularly
striking, when coming from the low $B$ region into the pinning
region. Indeed, at a fixed $B=4.3$T and $I=1mA$, the sample is
superconducting, then we can either increase the current to 2mA
and back or sweep $B$ to 4.5T and back, or increase $T$ to 1K and
back, and the sample is no longer superconducting and remains
dissipative on time scales beyond our experiment. Sweeping $B$ to
3.5T and back recovers a highly stable superconducting phase.
Hence, for the same $B$ and $I$ we can have two different phases,
which are stable over very long time scales. Moreover, when
increasing the temperature, the size of the region of hysteresis
decreases and eventually vanishes as illustrated in figure 4 (the
light blue shaded region). Since the dependence on temperature is
non-critical, this transition is strongly suggestive of a first
order phase transition inside the region of hysteresis, which we
indicated as a thick dashed line in figure 2. This is in agreement
with earlier experiments on crystals \cite{zeldov,twophases}, but
in our system this first order transition is not associated with
the formation of the {\em pinning} region but rather with the
transition to the {\em depinning 2} phase at $B$-fields above the
$PE$.

This additional {\em depinning 2} phase between the $PE$ and the
normal phase is only seen at sufficiently high driving currents
(figure 2) and is masked by the hysteresis at low $T$ and becomes
more apparent at higher $T$ as seen in figure 4. This hints to a
much richer transition region between the $PE$ and the {\em
normal} state than previously expected and constitutes an
important new experimental finding for the theoretical
understanding of these systems.

At even higher currents the {\em depinning 2} region can be
reached directly from the {\em depinning 1} region, which
corresponds to a sudden delocalization of the vortices identified
by a jump in the resistance. This region is not an inhomogeneous
mix of normal and superconducting regions because at low currents
the superconducting - to - normal transition is extremely sharp,
i.e., less than 50mT wide for the 10-90\%, which excludes any
large scale inhomogeneities in the sample. This type of transition
is consistent with several theoretical and numerical results
\cite{giamarchiDyn,olson,fangohr} describing the transition from a
$MBG$ to a smectic or plastic flow of vortices. Moreover, the {\em
depinning 2} region is increasingly geometry dependent at large
currents above 2mA as observed from tilted field measurements,
which suggests that the vortex order of the {\em depinning 2}
region increases with current as expected from the above-mentioned
theoretical results.

However, there is an overall downward bending of the pinning
region with current, which implies that for some fields vortices
get pinned again before they eventually depin. This is very
unusual and has never been observed before in the context of
vortices. We analyzed this behavior in more detail with the help
of the $IV$'s shown in figure 3. For $B\simeq$4T there is a
negative differential resistance as a function of current, which
implies a dynamical pinning mechanism at high average vortex
velocities. This effect cannot be due to an increase of effective
$T$ since $B_{c2}$ is almost not affected by currents in this
range, but it is hysteretic and might be due to the non-monotonous
ordering as a function of driving force, similar to the situation
of driven charge-density waves \cite{grant}.

\begin{figure}[h]
\begin{center}
\vspace*{1.5cm}
\includegraphics[scale=0.35]{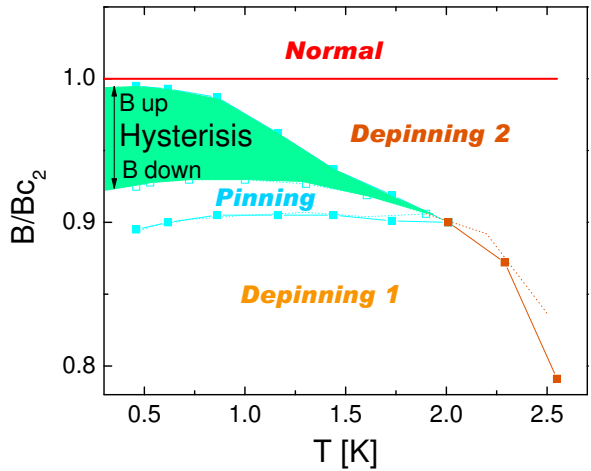}
\vspace*{-3cm} \caption{Phase diagram as a function of
temperature. The data was rescaled to $Bc_2$ and taken for $B$
applied parallel to the wide axis of the sample.}
\end{center}
\end{figure}

Summarizing, we have studied a new class of superconductors in
relation to the peak effect, in which the depinning threshold is
extremely weak and where there is no underlying crystalline order.
In our system the $PE$ is huge, with an increase by more than an
order of magnitude of the critical current in the $PE$ region,
indicating that very low pinning tends to enhance the peak effect.
In addition, after mapping out the entire phase diagram of our
system, while many similarities with previous studies on the $PE$
have been found, we also observed two striking differences.
Indeed, we observed an additional transition between the $PE$ and
the normal state as well as a dynamical pinning transition at
constant vortex density. These results have important implications
on our understanding of the dynamics of vortices in very weakly
pinned systems.

{\scriptsize {\bf Acknowledgements}

The authors would like to acknowledge helpful discussions with T.
Ala-Nissila, M. Grant, H. Guo and J.O. Strom-Olsen and support
from NSERC and FCAR.

\noindent (hilke@physics.mcgill.ca) }
\end{document}